\documentclass[prl,twocolumn,floatfix,showpacs,reprint,superscriptaddress]{revtex4-2}
\usepackage{color}
\usepackage{epsfig}
\usepackage{graphicx}
\usepackage{amsmath}
\usepackage[colorlinks, linkcolor=blue, citecolor=blue, urlcolor=blue]{hyper ref}

\begin{document}

\title{Nature of nonanalytic chemical short-range order in metallic alloys}

\author{Hao Deng}
\affiliation{School of Materials Science and Engineering, Northwestern Polytechnical University, Xi'an 710072, China}
\affiliation{China National Erzhong Group Deyang Wanhang Die Forging Co., Ltd, Deyang, 618000, China}

\author{Jue-Yi Qi}
\affiliation{School of Materials Science and Engineering, Northwestern Polytechnical University, Xi'an 710072, China}

\author{Qin-Han Xia}
\affiliation{School of Materials Science and Engineering, Northwestern Polytechnical University, Xi'an 710072, China}
\affiliation{State Key Laboratory of Solidification Processing, Northwestern Polytechnical University, Xi’an 710072, China}

\author{Jinshan Li}
\email{ljsh@nwpu.edu.cn}
\affiliation{School of Materials Science and Engineering, Northwestern Polytechnical University, Xi'an 710072, China}
\affiliation{State Key Laboratory of Solidification Processing, Northwestern Polytechnical University, Xi’an 710072, China}

\author{Xie Zhang}
\email{xie.zhang@nwpu.edu.cn}
\affiliation{School of Materials Science and Engineering, Northwestern Polytechnical University, Xi'an 710072, China}
\affiliation{State Key Laboratory of Solidification Processing, Northwestern Polytechnical University, Xi’an 710072, China}

\begin{abstract}
Nonanalytic chemical short-range order (SRO) has long been observed in diffuse scattering experiments for metallic alloys. 
However, considerable debate surrounds the validity of these observations due to the unresolved nature of the nonanalyticity.
Using prototypical face-centered cubic alloys as an example, here we demonstrate that SRO in metallic alloys is mostly nonanalytic at $\Gamma$.
The nonanalyticity stems from the elastic anisotropy and long-range atomic interactions of the \emph{host} lattice.
The physical insights substantially improve our understanding of chemical order in alloys and resolves the long-standing debate in the field.
Nonanalytic SRO is expected to be general in alloys and the nonanalyticity may serve as a unique feature to verify the intensely debated existence of SRO in compositionally complex alloys.
\end{abstract}

\maketitle

Within a fixed chemical composition, the physical properties of an alloy can be tuned by its atomic ordering, which is mediated by a competition between the atomic interactions and configurational entropy.
When the configurational entropy dominates, a chemically disordered alloy is usually formed.
However, if the atomic interactions govern, long-range order or phase separation may occur.
For an intermediate situation, the atomic interactions may give rise to local order, forming energetically preferred atomic pairs, i.e., chemical short-range order (SRO).

A fully random alloy is characterized by a uniform Laue scattering.
The presence of SRO modulates the Laue intensity, leading to diffuse peaks between or close to the Bragg reflections.
For a binary alloy ${\rm A}_{1-x}{\rm B}_{x}$, the diffuse SRO scattering intensity ($I_{\rm SRO}$) is directly proportional to the Warren-Cowley SRO parameter ($\alpha_{\bf q}$) in reciprocal space, i.e.,~\cite{krivoglaz_x-ray_1995,khachaturyan_theory_1983}
\begin{equation}
I_{\rm SRO}  \propto \alpha_{\bf q} = \frac{1}{1+\frac{x(1-x)}{k_{\rm B}T}V_{\bf q}^{\rm tot}}  ,
\end{equation}
where $k_{\rm B}$ is the Boltzmann constant and $T$ is temperature. 
$V_{\bf q}^{\rm tot}$ is the total atomic interaction, consisting of chemical ($V_{\bf q}^{\rm ch}$) and strain-induced ($V_{\bf q}^{\rm si}$) interactions, i.e., $V_{\bf q}^{\rm tot} = V_{\bf q}^{\rm ch} + V_{\bf q}^{\rm si}$.

In diffuse x-ray diffraction experiments for metallic alloys, a nonanalytic behavior for the diffuse scattering intensity at the $\Gamma$ point ($|{\bf q}| = 0$) in reciprocal space was observed as a generic feature in the Cu$_{83}$Mn$_{17}$ spin-glass alloy~\cite{reichert_strain-induced_2001,reichert_reichert_2002}, i.e., the diffuse scattering intensity is different while approaching $\Gamma$ from different directions.
This was also observed later in Cu-Au~\cite{reichert_absence_2003}, Au-Ni~\cite{reichert_competition_2005},  Ti-V~\cite{ramsteiner_omega-like_2008}, and Mg-Al~\cite{kurta_long-wavelength_2010} alloys.
The identification of the nonanalyticity is of critical importance, since it intimately relates to the SRO that impacts the physical properties of alloys.

However, the observed nonanalyticity was quickly challenged by theoretical calculations~\cite{chepulskii_comment_2002}, and neutron scattering experiments~\cite{schonfeld_analytical_2002}, in which no nonanalyticity was found. 
It is actually difficult to reach a solid conclusion, since the scattering experiments cannot directly access the $\Gamma$ point, but an extrapolation to $\Gamma$ has to be employed. 
For the theoretical calculations in Ref.~\cite{chepulskii_comment_2002}, the microscopic elasticity theory (MET) with empirical parametrization was used to compute the strain-induced interactions.
It was possible that the observed (non)analyticity is an artifact that arises from a certain parametrization of the key ingredients to the MET.

The chemical interaction is short-range and cannot introduce any nonanalyticity. 
Hence, the nonanalyticity in the diffuse scattering (if exists) should stem from the long-range strain-induced interaction.
Here we therefore employ an \textit{ab initio}-based MET approach to compute the strain-induced interactions for a series of typical face-centered cubic (fcc) alloys. 
We find that the nonanalyticity has a valid physical cause.
Aided by a rigorous theoretical derivation, we demonstrate that the nonanalyticity primarily originates from the elastic anisotropy of the host lattice. 
For Pd, Pt, and Pb based fcc alloys long-range force constants further contribute to the nonanalyticity.
Our important insights into the nature of the nonanalyticity of SRO not only resolves the debate on this experimental observation, but may also serve as a unique feature to assess the existence of  SRO in compositionally complex alloys.

In reciprocal space the strain-induced interaction can be computed using the MET~\cite{khachaturyan_theory_1983}
\begin{equation}
\label{eq:met}
V^{\rm si}({\bf q}) = -{\bf F}({\bf q}){\bf G}({\bf q}){\bf F}^{*}({\bf q}) + Q ,
\end{equation}
where ${\bf F}$ is the Kanzaki forces on the host lattice caused by the solute atom and ${\bf G}$ is the lattice Green's function. 
$Q$ is the self-interaction correction, $Q = 1/N\sum_{\bf q}{\bf F}({\bf q}){\bf G}({\bf q}){\bf F}^{*}({\bf q})$. 
$N$ is the total number of $q$-points used for sampling the first Brillouin zone. 
Instead of approximating ${\bf F}$ and ${\bf G}$ with lattice constants, elastic constants, and lattice distortion parameters, here we rigorously compute both ${\bf F}$ and ${\bf G}$ from \textit{ab initio} for dilute alloys that are easier to interpret.

\begin{figure}[ht]
\includegraphics[angle=0,width=75mm]{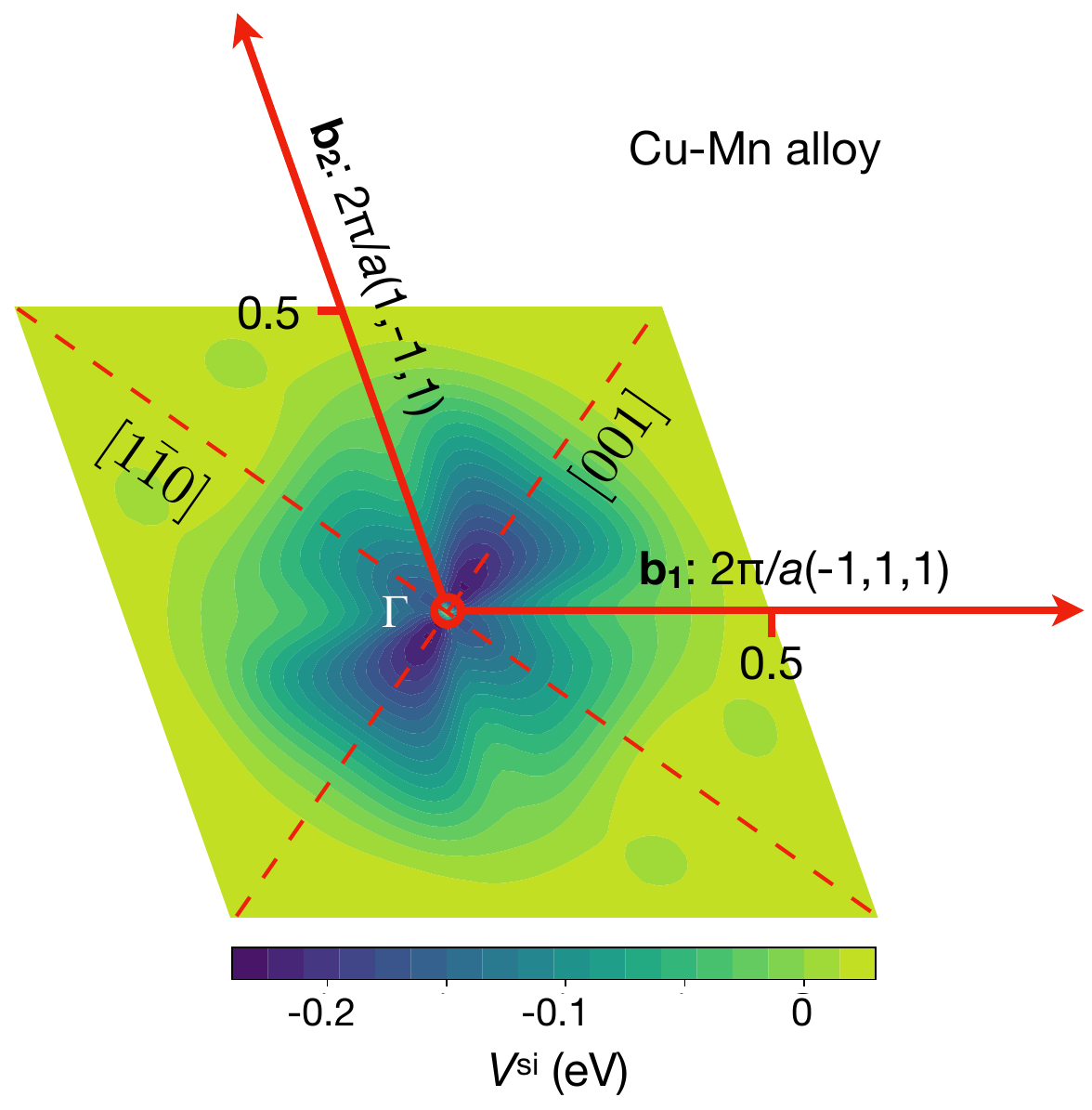}
\caption{\label{fig:SI-heatmap}
Strain-induced interaction $V^{\rm si}$ between Mn atoms in an fcc Cu host lattice within the ({${\bf b_1}$, ${\bf b_2}$}) plane in reciprocal space. ${\bf b_1}$ and ${\bf b_2}$ are two reciprocal lattice vectors.}
\end{figure}

\begin{figure*}[ht]
\includegraphics[angle=0,width=170mm]{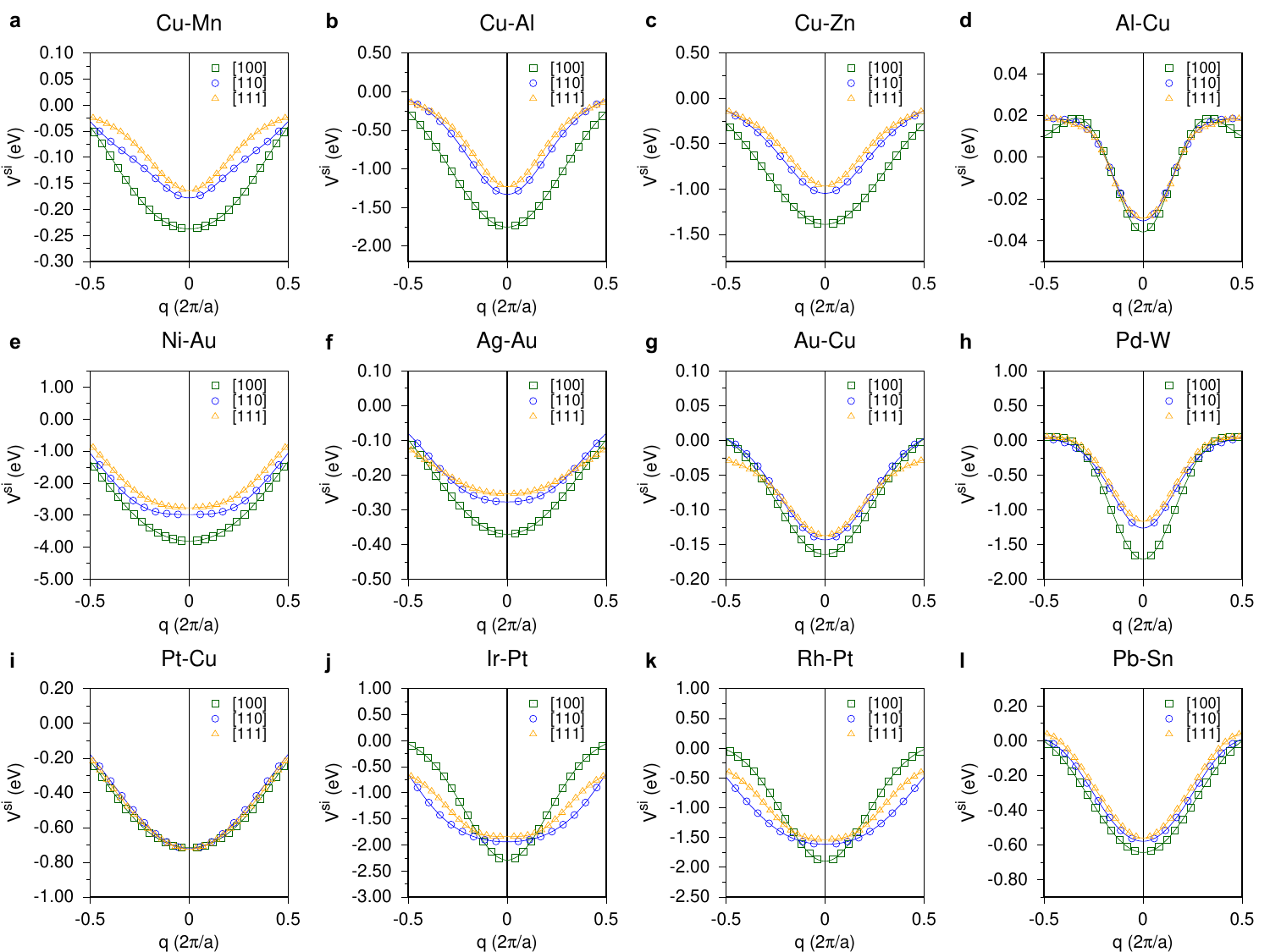}
\caption{\label{fig:SI-lineprofile}
Line profiles of the strain-induced interaction between solute atoms in a series of fcc alloys along three high-symmetry directions ([100], [110], and [111]) in reciprocal space. 
Here we consider binary alloys A$_{1-x}$B$_x$ with $x$ being very small, i.e., A is the host and B is the solute with dilute concentrations.}
\end{figure*}

Figure~\ref{fig:SI-heatmap} shows the strain-induced interaction between substitutional Mn atoms in the prototypical Cu-Mn alloy within the (${\bf b_1}$, ${\bf b_2}$) plane, where ${\bf b_1}$ and ${\bf b_2}$ are two reciprocal lattice vectors. 
It can be seen that $V^{\rm si}$ is highly anisotropic in the vicinity of $\Gamma$. 
When approaching $\Gamma$ from different directions, $V^{\rm si} (\Gamma)$ thus has different values. 
To better illustrate the nonanalyticity we depict in Fig.~\ref{fig:SI-lineprofile}(a) the line profiles of $V^{\rm si}$ in Cu-Mn along three high-symmetry directions: [100], [110] and [111]. 
Even though there is no nonanalyticity when going from $-{\bf q}$ to $+{\bf q}$ for each direction, the actual value of $V^{\rm si}_{100} (\Gamma)$ is substantially lower than those of $V^{\rm si}_{110} (\Gamma)$ and $V^{\rm si}_{111} (\Gamma)$, indicating a clear nonanalyticity at $\Gamma$. 
This is consistent with previous reports by Reichert \textit{et al.}~\cite{reichert_strain-induced_2001,reichert_reichert_2002}. 
We note that with our dense $q$-point mesh robust extrapolations to $\Gamma$ can be made, which rules out a potential error induced by extrapolation.

To investigate the universality of the observed nonanalyticity we compute $V^{\rm si}$ for two other Cu-based alloys and nine other fcc alloys that cover all typical fcc host lattices.
The substitutional elements are selected based on their importance in technological applications. 
The corresponding line profiles are shown in Figs.~\ref{fig:SI-lineprofile}(b)-\ref{fig:SI-lineprofile}(l).
There are three important observations that can be made from the results. 
First, the three representative Cu-based alloys show qualitatively similar nonanalyticities [Figs.~\ref{fig:SI-lineprofile}(a)-\ref{fig:SI-lineprofile}(c)], indicating that the nonanalyticity may be dominated by the host lattice. 
Second, even though the values of $V^{\rm si}$ vary for different alloys, a nonanalyticity at $\Gamma$ exists for most of the alloys inspected. 
An exception is the Pt-Cu alloy [see Fig.~\ref{fig:SI-lineprofile}(i)], for which no nonanalyticity is observed in a large vicinity of $\Gamma$. 
Third, Al-Cu, Au-Cu, Ir-Pt, Rh-Pt, and Pb-Sn alloys exhibit much less pronounced nonanalyticities compared to the other ones. 

To understand the fundamental origin of the nonanalyticity and explain our observations for $V^{\rm si}$ in different alloys, a theoretical derivation of the nonanalyticity is extremely helpful.
For most fcc metals the dominant interactions are the ones between first nearest neighbors, which is also a basic assumption within the conventional MET. 
Hence, we likewise systematically derive $V^{\rm si}$ within the approximation of first nearest-neighbor interactions. 

If we assume that the real-space Kanzaki force on the first nearest-neighbor atom ($a/2$, $a/2$, 0) is ($f$, $f$, 0) (equivalently for the other 11 first nearest-neighbor atoms), then the Fourier components of the Kanzaki forces can be derived as 
\begin{equation}
\label{eq:kanzaki}
\begin{split}
&F_x({\bf q}) = 4if\sin\frac{q_xa}{2}(\cos\frac{q_ya}{2} + \cos\frac{q_za}{2} ),\\
&F_y({\bf q}) = 4if\sin\frac{q_ya}{2}(\cos\frac{q_za}{2} + \cos\frac{q_xa}{2} ),\\
&F_z({\bf q}) = 4if\sin\frac{q_za}{2}(\cos\frac{q_xa}{2} + \cos\frac{q_ya}{2} ),
\end{split}
\end{equation}
where $a$ is the lattice constant, and $i$ is the imaginary unit.
The dynamical matrix ${\bf D}({\bf q})$ of an fcc lattice considering only first nearest-neighbor interactions reads in reciprocal space~\cite{krivoglaz_x-ray_1995}
\begin{equation}
\label{eq:dynmat}
\begin{split}
D_{11}({\bf q}) =& ac_{11}\left[ 2 - \cos\frac{q_xa}{2} (\cos\frac{q_ya}{2} + \cos\frac{q_za}{2}) \right] \\
&+ a(2c_{44} - c_{11})(1-\cos\frac{q_ya}{2} \cos\frac{q_za}{2}),\\
D_{12}({\bf q}) =& a(c_{12} + c_{44})\sin \frac{q_xa}{2} \sin\frac{q_ya}{2},
\end{split}
\end{equation}
where $c_{11}$, $c_{12}$, and $c_{44}$ are the elastic constants of a cubic crystal. The other seven components of the dynamical matrix can be obtained by cyclic permutation.

Within the above approximation deriving the nonanalyticity at $\Gamma$ is equivalent to computing the limit of $V^{\rm si}$ at $\Gamma$ when approaching from different directions. We introduce two parameters $\alpha$ and $\beta$ to define a direction by the vector [1, $\alpha$, $\beta$], i.e., $q_y = \alpha q_x$ and $q_z = \beta q_x$. 
Combining Eqs.~\eqref{eq:met}-\eqref{eq:dynmat} we can then rigorously derive $\lim_{{\bf q}\to0} V^{\rm si}$ as 

\begin{widetext}
\begin{equation}
\label{eq:vsqlimit}
\lim_{{\bf q}\to0} V^{\rm si} =\frac{-64f^2[12(\frac{1}{\eta}-1)^2\alpha^2\beta^2 + 4(\frac{1}{\eta}-1)(1+\alpha^2+\beta^2)(\alpha^2\beta^2+\alpha^2+\beta^2)+(1+\alpha^2+\beta^2)^3]}{4a(\frac{1}{\eta}-1)^2(c_{11}+2c_{12} +c_{44})\alpha^2\beta^2 + 2a(\frac{1}{\eta}-1)(c_{11} + c_{12}) (1+\alpha^2+\beta^2)(\alpha^2\beta^2+\alpha^2+\beta^2)+ ac_{11}(1+\alpha^2+\beta^2)^3} 
\end{equation}
\end{widetext}
Our detailed derivation of Eq.~\eqref{eq:vsqlimit} is provided in the Supplemental Material~\footnote{See Supplemental Materials at [URL] for details for the \textit{ab initio} calculations and the analytical derivations, which includes Refs.~\cite{kresse_efficient_1996,blochl_projector_1994,perdew_generalized_1996,monkhorst_special_1976,methfessel_high-precision_1989,grabowski_ab_2007,boeck_object-oriented_2011,murnaghan_compressibility_1944}.}.
Here, $\eta$ is the elastic anisotropy index as defined by Zener, i.e., $\eta = 2c_{44}/(c_{11}-c_{12})$~\cite{zener_elasticity_1948}. 

\begin{figure*}[ht]
\includegraphics[angle=0,width=180mm]{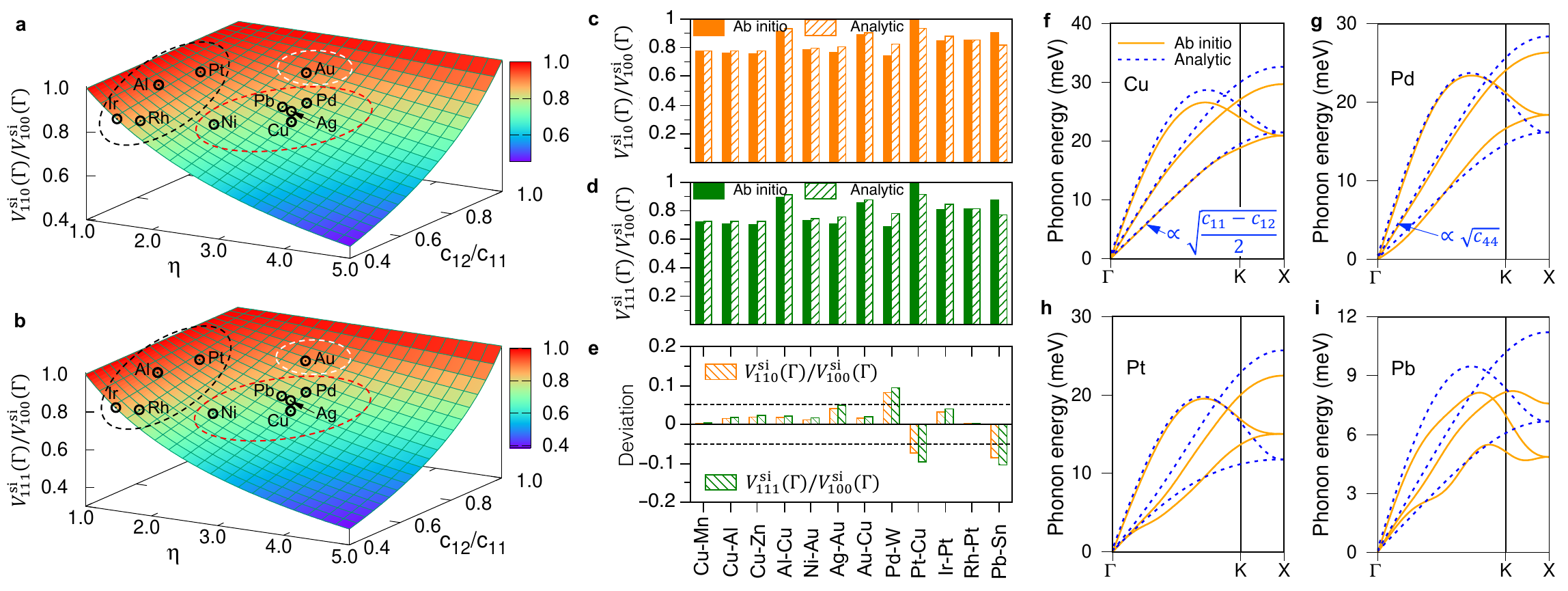}
\caption{\label{fig:category-n-phonon}
(a) $V^{\rm si}_{110}(\Gamma)/V^{\rm si}_{100}(\Gamma)$ and (b) $V^{\rm si}_{111}(\Gamma)/V^{\rm si}_{100}(\Gamma)$ as a function of the elastic anisotropy $\eta$ and $c_{12}/c_{11}$. (c, d) Comparison and (e) deviation between \textit{ab initio} computed and theoretically derived $V^{\rm si}_{110}(\Gamma)/V^{\rm si}_{100}(\Gamma)$  and $V^{\rm si}_{111}(\Gamma)/V^{\rm si}_{100}(\Gamma)$ for different fcc alloys. (f-i) Phonon dispersions obtained from \textit{ab initio} calculations and analytic formulas for Cu, Pd, Pt and Pb, respectively.}
\end{figure*}

The two ratios, $V^{\rm si}_{110} (\Gamma)/V^{\rm si}_{100} (\Gamma)$ and $V^{\rm si}_{111} (\Gamma)/V^{\rm si}_{100} (\Gamma)$, that characterize the nonanalyticity can be straightforwardly calculated by substituting the corresponding values of $\alpha$ and $\beta$ in  Eq.~\eqref{eq:vsqlimit}:
\begin{widetext}
\begin{equation}
\label{eq:analytic}
\begin{split}
\frac{V^{\rm si}_{110} (\Gamma)}{V^{\rm si}_{100} (\Gamma)} =& \frac{2}{1+\eta + (1-\eta)c_{12}/c_{11}},\\
\frac{V^{\rm si}_{111} (\Gamma)}{V^{\rm si}_{100} (\Gamma)} = &\frac{3(2/\eta +1)^2}{4(1/\eta-1)^2[1+2c_{12}/c_{11} +\eta/2(1-c_{12}/c_{11})] + 18(1/\eta-1)(1+c_{12}/c_{11}) + 27}.
\end{split}
\end{equation}
\end{widetext}

Based on our theoretical derivation it becomes clear that the nonanalyticity is mainly determined by the elastic properties of the host lattice,  $\eta$ and $c_{12}/c_{11}$. This is consistent with our \textit{ab initio} results that different Cu-based alloys show almost identical ratios of the strain-induced interaction approaching $\Gamma$ from different directions [Figs.~\ref{fig:SI-lineprofile}(a)-\ref{fig:SI-lineprofile}(c)].
We have also performed \emph{ab initio} calculations of $V^{\rm si}$ for several other fcc alloys with the same host lattice. Similar as for Cu-based alloys the nonanalyticity depends only on the host lattice.
As two limiting cases, when $\eta = 1$ or $c_{12}/c_{11}=1$,  both $V^{\rm si}_{110} (\Gamma)/V^{\rm si}_{100} (\Gamma)$ and $V^{\rm si}_{111} (\Gamma)/V^{\rm si}_{100} (\Gamma)$ are equal to 1. 
In these two cases $V^{\rm si} (\Gamma)$ is direction-independent with $V^{\rm si} (\Gamma) = -64f^2/(ac_{11})$, thus exhibiting an analytic behavior. 
For other cases, there are nonanalyticities.
Notably, since $\lim_{\bf q\to0}V^{\rm si}$ is an even function of ${\bf q}$, it is analytic when going from $-{\bf q}$ to $+{\bf q}$, consistent with the \textit{ab initio} results in Fig.~\ref{fig:SI-lineprofile}. 

Using Eq.~\eqref{eq:analytic} we can investigate the overall chemical trends for the nonanalyticity by examining the dependence of the ratios on $\eta$ and $c_{12}/c_{11}$.
In Figs.~\ref{fig:category-n-phonon}(a) and~\ref{fig:category-n-phonon}(b) $V^{\rm si}_{110} (\Gamma)/V^{\rm si}_{100} (\Gamma)$  and $V^{\rm si}_{111} (\Gamma)/V^{\rm si}_{100} (\Gamma)$ are shown as a function of $\eta$ and $c_{12}/c_{11}$. 
The data points indicate the corresponding values  obtained from our \textit{ab initio} elastic constants to determine $\eta$ and $c_{12}/c_{11}$ (see Ref.~\cite{Note1} for details).
Based on the analytic results, all fcc host lattices that we inspected can be categorized into three groups [dashed ellipses in Figs.~\ref{fig:category-n-phonon}(a) and~\ref{fig:category-n-phonon}(b)]. The first group includes Al, Pt, Ir and Rh. 
The nonanalyticity is small in these alloys because of their low elastic anisotropy ($\eta$). 
The second group consists of Au. The elastic anisotropy of Au is not that low, but the $c_{12}/c_{11}$ ratio is relatively large, leading to a small nonanalyticity at $\Gamma$ as well. 
The third group is composed of Ni, Cu, Ag, Pb, and Pd. 
These metals exhibit a  relatively high elastic anisotropy and low $c_{12}/c_{11}$ value, resulting in pronounced nonanalyticities of $V^{\rm si}$.

This categorization is consistent with the trend observed for most of the alloys shown in Fig.~\ref{fig:SI-lineprofile}, but fails to explain, e.g., why the Pt-based alloy has an analytic behavior [Fig.~\ref{fig:SI-lineprofile}(i)]. 
Figures~\ref{fig:category-n-phonon}(c)-\ref{fig:category-n-phonon}(e) show a quantitative comparison of the nonanalyticity from direct \textit{ab initio} calculations and the theoretically derived formula [Eq.~\eqref{eq:analytic}]. 
Overall, we observe good agreement between the two approaches with deviations within $\pm$0.05 [dashed lines in Fig.~\ref{fig:category-n-phonon}(e)]. 
Larger discrepancies exist for Pd-W, Pt-Cu, and Pb-Sn alloys. 
Especially, for the Pt-based alloy, \textit{ab initio} calculations show analyticity at $\Gamma$, but the theoretical derivation predicts a noticeable nonanalyticity. 

As discussed, the nonanalyticity is determined by the elastic properties of the host lattice. Therefore, the above discrepancies may originate from elastic anomalies of the host lattice in the vicinity of $\Gamma$.
We therefore compute the phonon dispersions of all the fcc metals we inspected from both \textit{ab initio} calculations and the analytic formula in Eq.~\eqref{eq:dynmat}. 
Representatively, we show the phonon dispersions of Cu, Pd, Pt, and Pb in Figs.~\ref{fig:category-n-phonon}(f)-\ref{fig:category-n-phonon}(i). 

As a typical fcc metal that can be well described by the analytic model, the phonon dispersions of Cu using the two approaches shown in Fig.~\ref{fig:category-n-phonon}(f) are consistent in the vicinity of $\Gamma$. 
Similarly, we observe good agreement between the \emph{ab initio} and analytic phonon dispersions for many other fcc host lattices.
However, pronounced discrepancies between the two approaches are revealed for three metals: Pd, Pt, and Pb.
 In particular, the lowest two acoustic branches show sizable deviations. 
The slopes of the lowest two acoustic branches while approaching $\Gamma$ along the [110] direction are proportional to $\sqrt{(c_{11}-c_{12})/2}$ and $\sqrt{c_{44}}$, respectively. 
One can clearly see that for Pt and Pb the lowest two acoustic branches become almost degenerate when approaching $\Gamma$. 
A degeneracy of these two branches leads to an ``effective" $\eta$ of 1. 
This indicates that even though these two metals are not elastically isotropic, long-range atomic interactions within the host lattice counteract the elastic anisotropy, which thus leads to a pronounced decrease of the nonanalyticity at $\Gamma$. 
As for Pd, the long-range interactions also play an important role, but in contrast to Pt, further enhance the elastic anisotropy, resulting in an increase of the nonanalyticity at $\Gamma$. 
These insights well explain the differences between the \emph{ab initio} and theoretically derived results in  Figs.~\ref{fig:category-n-phonon}(c) and~\ref{fig:category-n-phonon}(d).

\begin{figure}[ht]
\includegraphics[angle=0,width=85mm]{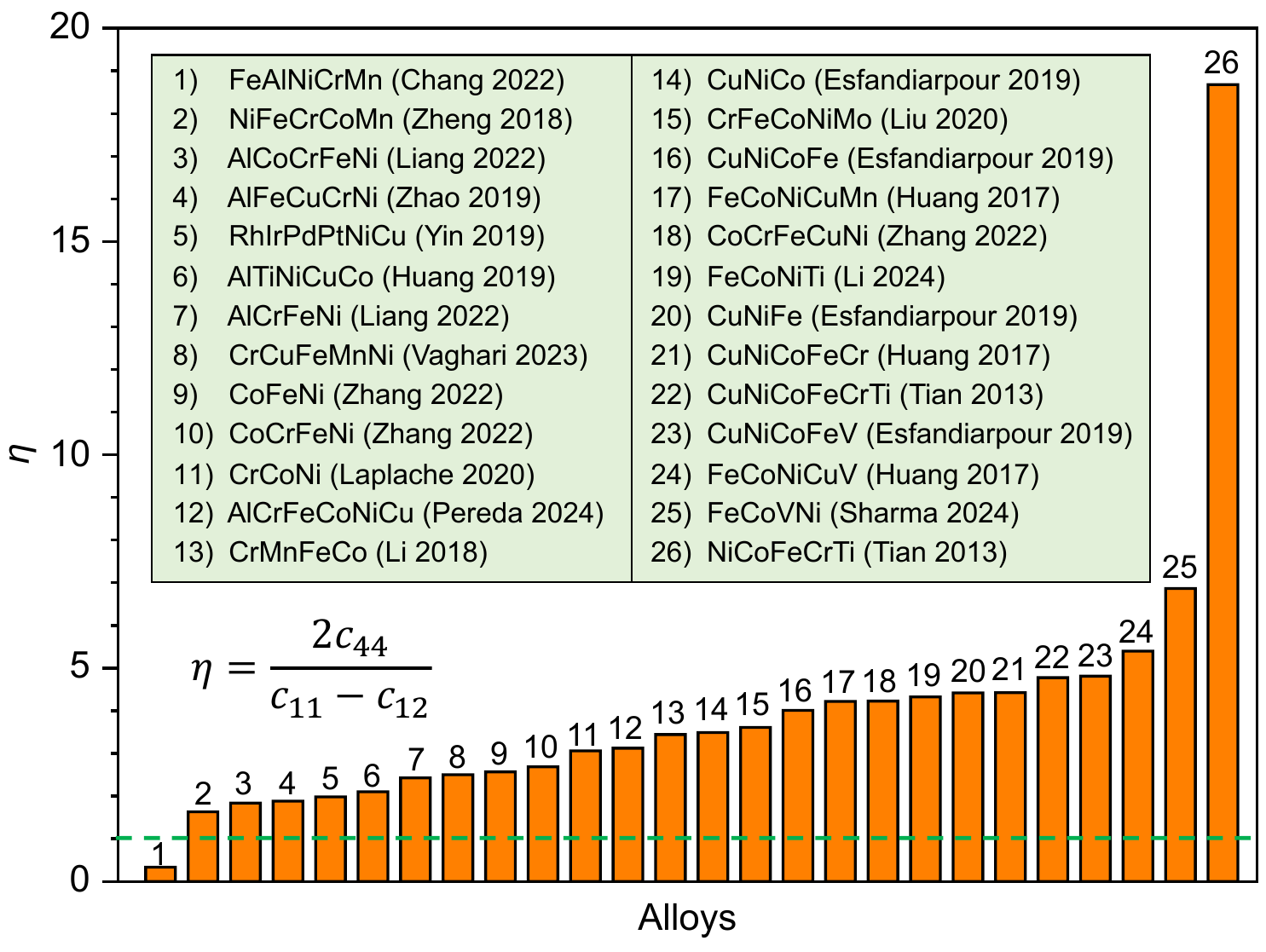}
\caption{\label{fig:eta}
Computed elastic anisotropy indices for various fcc alloys with their elastic constants extracted from the literature~\cite{chang_first-principles_2022,zheng_elastic_2018,liang_phase_2022,zhao_effect_2019,yin_first-principles-based_2019,huang_evaluating_2019,vaghari_computational_2023,zhang_predicting_2022,laplanche_processing_2020,pais_pereda_analysis_2024,li_third-order_2018,esfandiarpour_investigation_2019,liu_theoretical_2020,huang_thermal_2017,li_designing_2024,tian_ab_2013,sharma_first-principles_2024}.}
\end{figure}

Our detailed analysis above demonstrates that the nonanalyticity in the diffuse scattering at $\Gamma$ primarily originates from the elastic anisotropy of the host lattice.
This insight can actually be very useful for resolving the recent controversial reports for the SRO in compositionally complex alloys~\cite{zhang_short-range_2020,chen_direct_2021,ferrari_simulating_2023,moniri_three-dimensional_2023,yin_yield_2020,pei_rigorous_2025}.
It has been shown that other sources such as planar defects and higher-order Laue zone reflections may also give rise to similar extra electron reflections~\cite{walsh_extra_2023,coury_origin_2023}, making the experimental analysis of SRO difficult.

The nonanalyicity in diffuse scattering revealed in this study focuses on SRO, which is captured at the long-wavelength limit ($|{\bf q}| \approx 0$).
In Fig.~\ref{fig:eta} we examine the elastic anisotropy ($\eta$) for various compositionally complex fcc alloys using their elastic constants from the literature.
One can see that these alloys are all elastically anisotropic, which means that nonanalytic diffuse scattering is expected to be observable in these alloys.
Diffuse scattering experiments with high-energy beams (preferably synchrotron sources that allow for access to as small {\bf q} as possible) would enable rigorous experimental assessment of the  SRO in compositionally complex alloys.

To conclude, we demonstrate that the nonanalyticity at $\Gamma$ for diffuse x-ray scattering is a generic feature in fcc alloys. 
The nonanalyticity is fundamentally mediated by  SRO in an elastic anisotropic host lattice. 
The long-range interactions in Pd-, Pt-, and Pb-based host lattices may further impact the amplitude of the nonanalyticity.
The nonanalyticity of SRO in diffuse scattering revealed here may serve as a unique feature to distinguish between SRO and other sources.

\begin{acknowledgments}
The authors are grateful to Dr. Tilmann Hickel, Dr. Jutta Rogal, and Prof. J\"{o}rg Neugebauer for fruitful discussions.
This work was supported by National Major Science and Technology Projects of China (2024ZD0608100).
The authors acknowledge the Beijing Super Cloud Center (BSCC) for providing HPC resources that have contributed to the research results reported within this paper (URL: http://www.blsc.cn/).
\end{acknowledgments}

\end{document}